\documentclass[12pt]{elsart}
\usepackage{hyperref}
\usepackage{epsfig}
\usepackage{setspace}
\usepackage{longtable}

\begin{document}
\begin{frontmatter}
\title{Kink plateau dynamics in finite-size lubricant chains}
\author{Marco Cesaratto$^{a}$},
\author{Nicola Manini$^{a}$},
\author{Andrea Vanossi$^{b}$},
\author{Erio Tosatti$^{c,d}$, and}
\author{Giuseppe E. Santoro$^{c,d}$\corauthref{cor}}
\ead{santoro@sissa.it} \corauth[cor]{Corresponding Author.}
\address{$^a$Department of Physics, University of Milan, Via Celoria 16, 20133 Milan, Italy}
\address{$^b$CNR-INFM National Research Center S3, and Department of Physics,
University of Modena and Reggio Emilia, Via Campi 213/A, 41100
Modena, Italy}
\address{$^c$International School for Advanced Studies (SISSA),
and INFM-CNR Democritos National Simulation Center, Via Beirut 2-4,
I-34014 Trieste, Italy}
\address{$^d$International Centre for
Theoretical Physics (ICTP), P.O.Box 586, I-34014 Trieste, Italy}


%
\maketitle
\begin{abstract}
We extend the study of velocity quantization phenomena recently
found in the classical motion of an idealized 1D model solid
lubricant -- consisting of a harmonic chain interposed between two
periodic sliding potentials [Phys.\ Rev.\ Lett.\ {\bf 97}, 056101
(2006)]. This quantization is due to one slider rigidly dragging
the {\em commensurate} lattice of kinks that the chain forms with
the other slider. In this follow-up work we consider finite-size
chains rather than infinite chains.
The finite size (i) permits the development of robust velocity
plateaus as a function of the lubricant stiffness, and (ii) allows
an overall chain-length re-adjustment which spontaneously promotes
single-particle {\em periodic} oscillations.
These periodic oscillations replace the quasi-periodic motion
produced by general incommensurate periods of the sliders and the lubricant
in the infinite-size model.
Possible consequences of these results for some real systems are discussed.
\end{abstract}

\end{frontmatter}

\section{Introduction}

The present paper extends the study of a one-dimensional (1D) non-linear
model, inspired by the tribological problem of two sliding surfaces with a
thin solid lubricant layer in between \cite{Vanossi06}, to the case of a
lubricant  1D ``island'' of finite size.
Previous work \cite{Vanossi06,Santoro06} found robust, universal
and exactly quantized asymmetric velocity plateaus in the
classical dynamics of an infinite-size chain subject to two
relatively sliding periodic potentials. The infinite chain size
was managed -- in the general incommensurate case --
 by means of periodic boundary conditions (PBC) and finite-size scaling.
The plateaus of chain velocity as a function of several model
parameters were shown to be due to motion of kinks (topological
solitons), which generally exist in a chain submitted to a
periodic potential. These nonlinear excitations generated by the
first, stationary periodic potential are set into motion by the
external driving which is provided by the second periodic
potential, sliding with velocity $v_{\rm ext}$. While the chain
kinks are thus dragged with velocity $v_{\rm ext}$, the overall
chain velocity is smaller, and fixed by the kinks nature and
density. That in turn depends only on the ratio of the period of
one slider to that of the chain, which is dictated by the
interparticle spacing, and enforced by the PBC.

The present work considers a finite open-boundary chain, such as
for example a hydrocarbon chain, or a graphite flake interposed
between two sliding crystal faces \cite{Dienwiebel04}.
Unlike the infinite chain, the open-boundary chain can elongate or shorten,
at the cost of some harmonic potential energy, effectively modifying its
linear density, and thus the kink density which is the relevant length
ratio. We find that it does indeed elongate or shorten so as to realize a
precise {\em commensurate} relation to the other slider.
This adaptive relaxation is such as to produce perfectly periodic oscillations of the
single particles superposed with their average drift at the quantized velocity.

\section{The model}

\begin{figure}
\centerline{
\epsfig{file=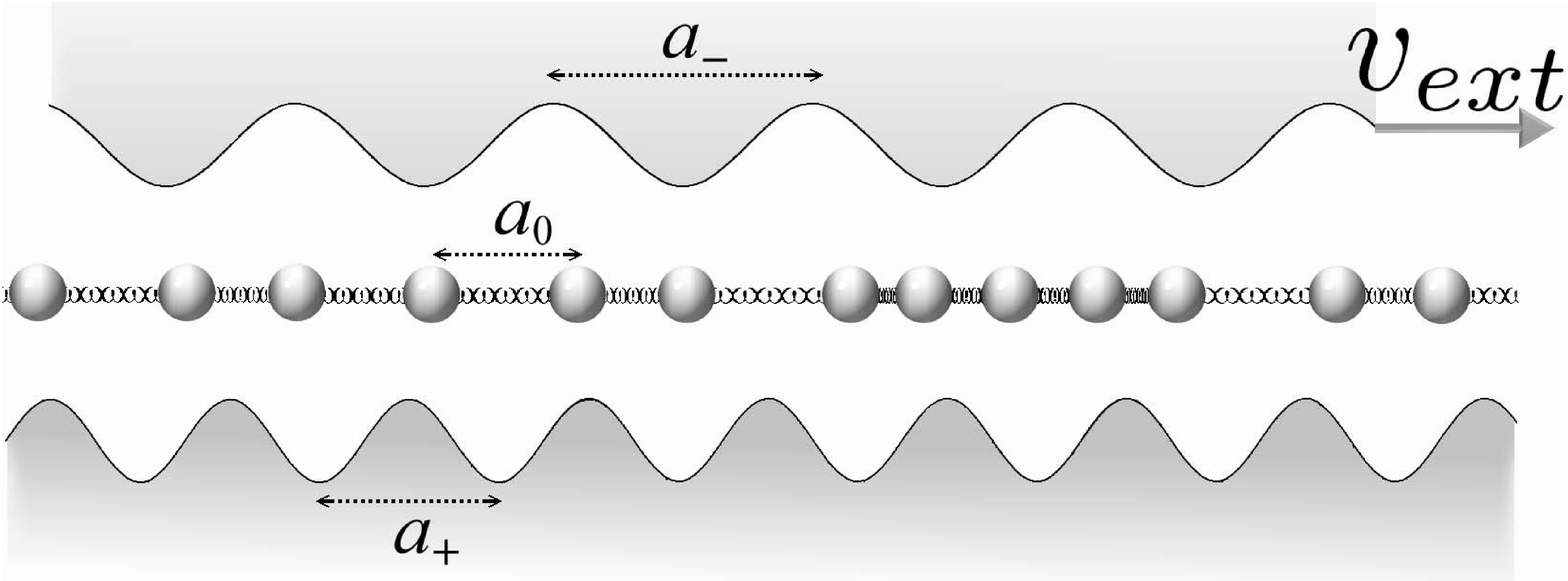,width=8cm,angle=0,clip=}
}
\caption{\label{model:fig}
(Color online) 
A sketch of the model with the two periodic sliders (periods $a_+$ and
$a_-$) and the lubricant harmonic chain of rest nearest-neighbor length
$a_0$.
}
\end{figure}

Using the same language and notation of previously studied
confined models
\cite{Rozman96,Rozman98,Zaloj98,Urbakh,VanossiPRL}, we represent a
solid lubricant layer as a chain of $N$ harmonically interacting
particles interposed between two rigid generally (but not
necessarily) incommensurate sinusoidal substrates (the two
``sliding crystals'', sketched in Fig.~\ref{model:fig}) externally
driven at a constant relative velocity $v_{\rm ext}$.
The equation of motion of the $i$-th particle is:
\begin{eqnarray} \label{eqmotion:eqn}
m\ddot{x}_i &=&  -\frac{1}{2} \left[ F_+ \sin{k_+ (x_i-v_+t)} + F_- \sin{k_-(x_i-v_-t)}\right] \nonumber \\
&+& K (x_{i+1}+x_{i-1}-2x_i) - \gamma \sum_{\pm} (\dot{x}_i - v_{\pm}) \;,
\end{eqnarray}
where $m$ is the mass of the $N$ particles, $K$ is the chain spring
constant, and $k_{\pm}=2\pi/a_{\pm}$ are the wave-vector periodicities of
potentials representing the two sliders, moving at velocities $v_{\pm}$.
We set, in full generality, $v_+$ = 0 and $v_- = v_{\rm ext}$.
$\gamma$ is a phenomenological parameter substituting for various sources
of dissipation, required to achieve a stationary state, but otherwise playing
no major role in the following.
$F_{\pm}$ are the force amplitudes representing the sinusoidal corrugation
of the two sliders (we will commonly use $F_-/F_+=1$ but we
checked that our results are more general).
We take $a_+=1$, $m=1$, and $F_+=1$ as our basic units, and all quantities
are measured in suitable combinations thereof.
The relevant length ratios \cite{vanErp99,Vanossi00} are defined by
$r_{\pm}=a_{\pm}/a_0$; we assume, without loss of generality, $r_->r_+$.
The inter-particle equilibrium length $a_0$ enters explicitly the
equations of motion (\ref{eqmotion:eqn}) of the first ($i=1$) and
last ($i=N$) particle whose restoring force terms in
Eq.~(\ref{eqmotion:eqn}) are $K (x_{2}-x_1-a_0)$ and $K
(x_{N-1}-x_{N}+a_0)$, respectively; this implements open boundary
conditions (OBC).

Upon sliding the substrates, $v_{\rm ext} \neq 0$, the lubricant
chain slides too. Despite the apparent generic symmetry between
the two sliders, the time-averaged chain velocity
$w=v_{\rm cm}/v_{\rm ext}$, is generally {\it asymmetric}, namely
different from $1/2$.
In a previous study on this model \cite{Vanossi06,Santoro06} it was
shown that, for an infinite chain and periodic boundary conditions (PBC), $w$
is exactly quantized, for large parameter intervals, to plateau values that depend
solely on the chosen commensurability ratios $(r_+,r_-)$.
As the present finite-size OBC simulations will show, the PBCs are
not crucial to the plateau quantization, which occurs even for a lubricant
of finite and not particularly large size $N$.
The main difference between OBC and PBC is that, while in PBC the chain
length, and thus the length ratios $r_\pm$, are fixed, in OBC the chain can
lengthen or shorten with respect to its equilibrium size.
We find that during sliding the chain length gradually reaches a natural
attractor value and oscillates around it.
We define a new effective inter-particle length and corresponding length ratio
\begin{equation}\label{r+eff}
a_0^{\rm eff}=\frac L{N-1}\,, \qquad r_+^{\rm eff}=\frac{a_+}{a_0^{\rm eff}}
\,,
\end{equation}
where $L=\langle x_{N}-x_{1}\rangle$ is the average chain length after the
initial transient.
The effective length ratio $r_+^{\rm eff}$ plays a central role in the
understanding of the velocity plateaus of the finite-size lubrication
model.

\section{Results and theory} \label{results:sec}

The driven dynamics of the lubricant is studied by integrating the
equations of motion (\ref{eqmotion:eqn}) starting from fully relaxed
springs ($x_i=i\, a_0$, $\dot{x}_i=v_{\rm ext}/2$), using a standard
fourth-order Runge-Kutta method.
After an initial transient where length relaxation takes place, the system
reaches its dynamical stationary state, at least so long as $\gamma$ is not
exactly zero.
Figure~\ref{w_K:fig} shows the resulting time-averaged center-of-mass (CM)
velocity $v_{\rm cm}$ as a function of the chain stiffness $K$, for an
irrational choice of $(r_+,r_-)$, and two values of $N$,
defining a relatively short chain ($N=15$) and one of intermediate length
($N=100$).
We find that $w$ is generally a complicated function of $K$, with flat
plateaus and regimes of continuous evolution, not unlike what is found in
the infinite-size limit studied through PBC simulations \cite{Vanossi06}.

\begin{figure}
\centerline{
\epsfig{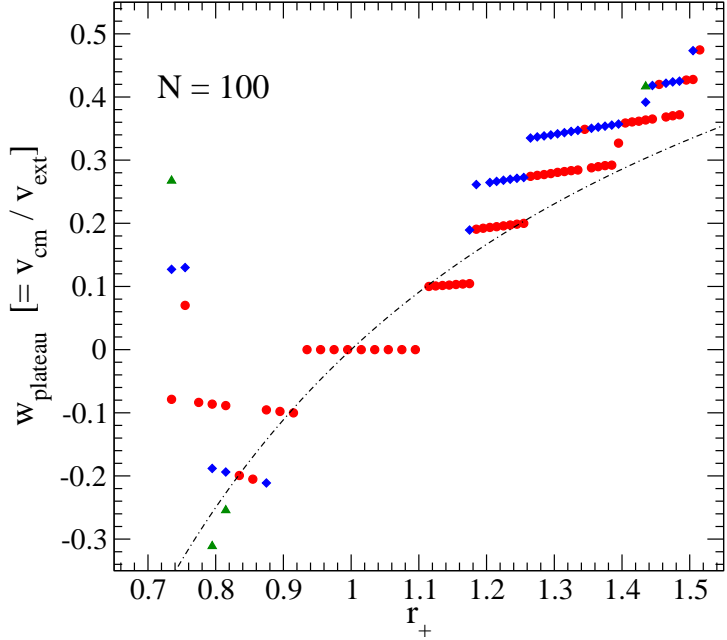}
}
\caption{\label{w_r:fig}
(Color online) 
The plateau velocity ratio $w$ as a function of $r_+$, for $r_-=101^{1/2}$.
Circles, diamonds and triangles represent respectively the first, second
and third plateau found for increasing $|w|$.
For comparison, we report the plateau velocity ratio
$w=1-r_+^{-1}$ (dot-dashed curve) of the fixed-length PBC calculations
\cite{Vanossi06}.
Here $\gamma=0.1$, $v_{\rm ext}=0.01$, and a finite chain of $N=100$ particles
(OBC) is considered.
}
\end{figure}

To investigate the peculiarities brought about by
finite size, we analyze the dynamics
for a large number of values of $(r_+,r_-)$ and $K$, and observe that:
(i) one or more velocity plateaus as a function of $K$ occur
for large ranges of $(r_+,r_-)$;
(ii) the velocity ratio $w$ of most
plateaus satisfies
\begin{equation}\label{weff}
w=1-\frac 1{r_+^{\rm eff}}
.
\end{equation}
This result should be compared with the relation ($w=1-r_+^{-1}$) valid for
the main plateau of PBC calculations \cite{Vanossi06}, where the
length ratios $(r_+,r_-)$ are fixed:
in Eq.~(\ref{weff}) the new effective length ratio $r_+^{\rm eff}$
of Eq.~(\ref{r+eff}) replaces $r_+$.
Figure~\ref{w_r:fig} collects the observed plateau velocity ratios for a
range of values of $r_+$ and for fixed $r_-=\sqrt{101}$.
Clear trends emerge: plateau velocities depend continuously on $r_+$, thus
indicating that also the effective chain length ratio $r_+^{\rm eff}$
evolves continuously with $r_+$ in finite ranges.
Several plateaus appear to follow different curves.

\begin{figure}
\centerline{
\epsfig{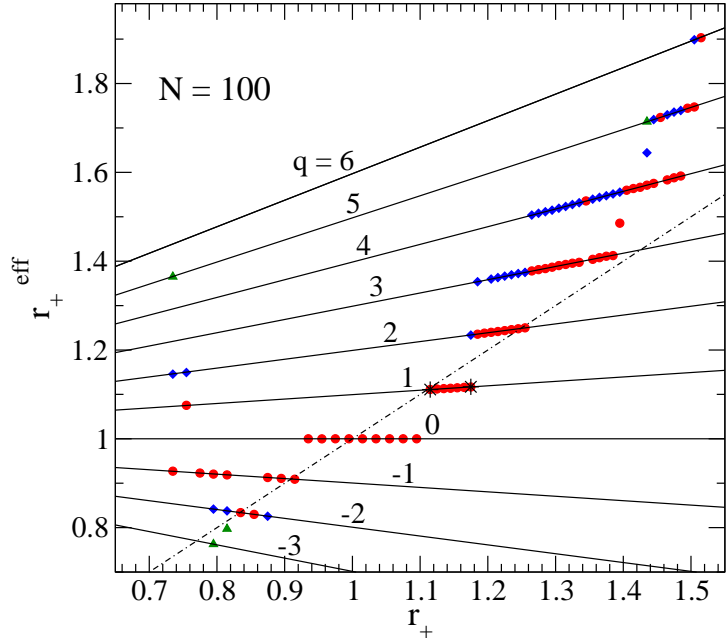}
}
\caption{\label{reff:fig}
(Color online) 
The effective length ratio $r_+^{\rm eff}=a_+/a_0^{\rm eff}$, calculated from the
velocity data of Fig. \ref{w_r:fig} using $r_+^{\rm eff}=(1-w)^{-1}$, the
inverse of Eq.~(\ref{weff}).
The stars are $r_+^{\rm eff}$ values obtained based on the definition
(\ref{reff:eq}) by time-averaging the chain length, showing perfect
agreement with the velocity-derived values.
The dot-dashed line represents the identity $r_+^{\rm eff}=r_+$ (undeformed
chain length).
Solid straight lines: $r_+^{\rm eff}=1+q \, \frac{r_+}{r_-}$, for integer
$q$.
}
\end{figure}

These data are conveniently organized and understood by plotting
$r_+^{\rm eff}$ rather than $w$, as a function of $r_+$, as is done in
Fig.~\ref{reff:fig}.\footnote{
For reasons of numerical stability, this figure reports $r_+^{\rm eff}$
obtained by inversion of Eq.~(\ref{weff}), but the same plot could have been
obtained directly by applying the definition (\ref{r+eff}). This
equivalence is confirmed by the comparison of the two methods shown in
Fig.~\ref{reff:fig} for 2 points.
}
All points fit perfect straight lines through $(r_+,r_+^{\rm eff})=(0,1)$.
Most lines have slope $q/r_-$ with integer $q$.
Occasional plateaus fit this relation with half-integer $q$ ($-5/2$, $7/2$,
and $9/2$, in the calculations of Fig.~\ref{reff:fig}).

Several calculations carried out with different values of $r_-$ confirm
that in general the ratio $r_+^{\rm eff}$ satisfies
\begin{equation} \label{reff:eq}
r_+^{\rm eff}=1+q \, \frac{r_+}{r_-}
\end{equation}
with $q$ taking simple fraction (often integer) values.
This general behavior indicates that the plateau dynamics leads the
finite-size lubricant toward a dynamical configuration where not only its
velocity but also its length is quantized.

We can understand this phenomenology in terms of kinks (i.e.,
local compressions of the chain with substrate potential minima
holding more than just one particle \cite{BraunBook}), as
described in \cite{Vanossi06}.
Assume initially integer $q$.
The basic hypothesis explaining the relation (\ref{reff:eq}) are:
(i) the particles tend to singly occupy the $a_+$-spaced minima of the
bottom potential, with occasional kinks to release the spring tension;
(ii) kinks group in bunches, each sitting in a period $a_-$ of the top
substrate;
(iii) kink bunches are $q$-fold, i.e.\ they collect $q$ individual kinks
(negative $q$ indicates the number of anti-kinks).
After the initial transient the chain length becomes on average very close
to $L = (N-N_{kink}-1)\,a_+$.
The number of kinks thus equals the number $q$ of kinks per bunch times the
total number $L/a_-$ of bunches in the chain: $N_{kink}= q\,L/a_-$.
By eliminating $N_{kink}$, we obtain
\begin{equation}
L=\frac{a_+a_-(N-1)}{a_-+q\,a_+}
\,.
\end{equation}
This is consistent with an average inter-particle distance
\begin{equation}
a_0^{\rm eff}=\frac{L}{N-1} = \frac{a_+a_-}{a_-+q\,a_+}
=a_+\,\frac{r_-}{r_-+q\,r_+}
=a_+\,\frac{1}{1+q\,\frac {r_+}{r_-}}
\,,
\end{equation}
and thus with the effective length ratio of Eq.~(\ref{reff:eq}).
In general, for non-integer $q=n_k/n_-$ values, this interpretation remains
valid: bunches of a total of $n_k$ kinks distribute themselves in $n_-$
minima of the $a_-$ lattice.
$q$ indicates therefore the density (coverage fraction) of kinks
on the $a_-$ lattice.


The dynamically stable plateau attractors of the open-boundary chain are
therefore characterized by a lattice of kinks perfectly commensurate to the
$a_-$ lattice.
In the infinite-size PBC model, a rational kink coverage corresponds to
commensurate encounter frequencies $f_+$ and $f_-$ of the generic lubricant
particle with the two substrates, which in turn occurs for very special values of
$r_\pm$ \cite{Vanossi06}: these $(r_+,r_-)$ values are 
characterized by perfectly periodic single-particle dynamics.
It is rather remarkable that the open-chain model, without the
necessity of any careful fine tuning of $(r_+,r_-)$ realizes a
self-organized commensurate kink lattice automatically producing perfectly
periodic single-particle oscillations in a generally incommensurate
context.

Note that the points of Fig.~\ref{reff:fig} lying along the $q=0$ line
indicate perfect commensuration between the lubricant chain and the bottom
substrate, to which the chain remains pinned: this can be realized by
paying a moderate harmonic-energy cost only close to $r_+=1$.
For the same reason, all plateaus appear close to the $r_+=r_+^{\rm eff}$ line
(dot-dashed line in Fig.~\ref{reff:fig}).

For the parameters of Fig.~\ref{w_r:fig}, nontrivial plateaus are found
only for $0.7<r_+< 1.6$.  Outside this range, relation (\ref{reff:eq})
produces $r_+^{\rm eff}$, for small integer $q$, very much different from $r_+$. 
Smaller values of $r_-\simeq 2$ (rather $r_-\simeq 10$ as in
Fig.~\ref{w_r:fig}) generates an analogous set of plateaus for $r_+> 1.6$.

\section{Size dependence}

\begin{figure}
\centerline{
\epsfig{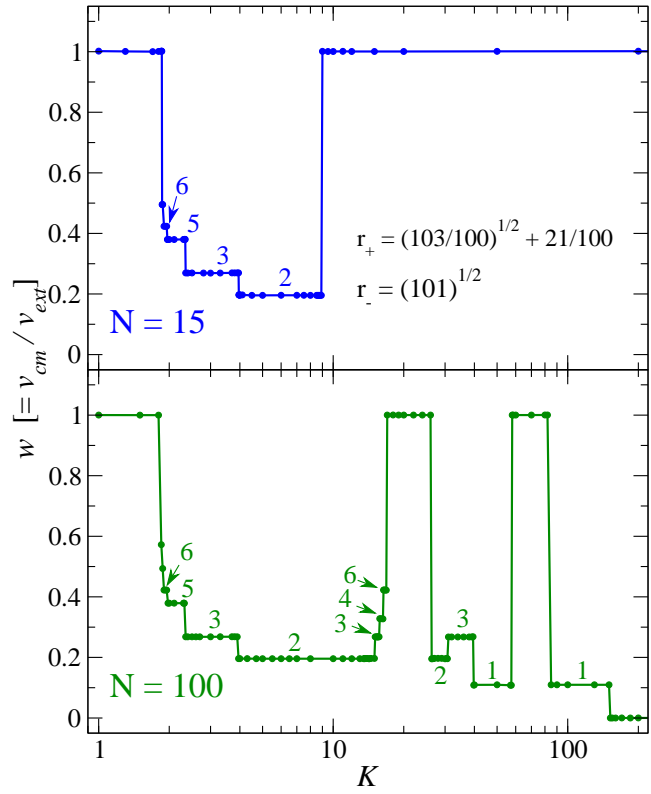}
}
\caption{\label{w_K:fig}
(Color online) 
Average drift velocity ratio $w=v_{\rm cm}/v_{\rm ext}$ as a function of the
lubricant chain spring stiffness $K$ for $(r_+,r_-)=
(\sqrt{\frac{103}{100}}+\frac{21}{100},\sqrt{101})\simeq (1.2249,10.050)$
and for two different finite sizes of the chain $N=15$ and $100$.
OBC are applied, and $\gamma=0.1$, $v_{\rm ext}=0.01$.
The plateau labeling refers to values of $q$, defined in
Eq.~(\ref{reff:eq}).
Note the similar behaviour for $K<10$ in spite of the different chain size,
and the different limit for large $K$, due to finite-size effects.
}
\end{figure}

\begin{figure}
\centerline{
\epsfig{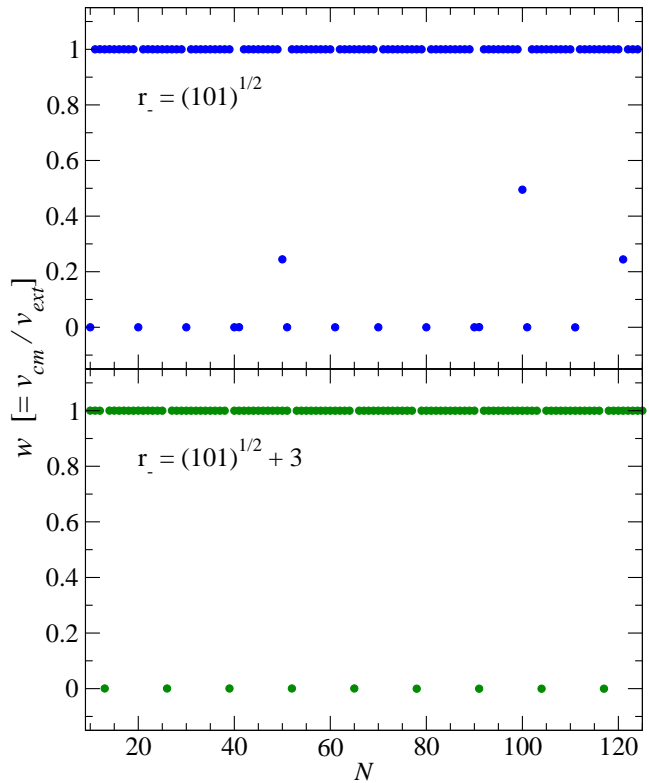}
}
\caption{\label{K1000:fig}
(Color online) 
Size scaling of the average drift velocity ratio $w$ for
$r_+=\frac{\sqrt{5}+1}{2}=\phi$, and two different values of the ratio
$r_-=(101)^{1/2}\simeq 10.05$ and $r_-=(101)^{1/2}+3\simeq 13.05$.
The number of particles $N$ for which the chain is pinned to the $a_+$
substrate ($w=0$) are close to integer multiples of $r_-$.
All calculations refer to $\gamma=0.1$, $v_{\rm ext}=0.01$, and $K=1000$, a
representative of the large-$K$ limit.
}
\end{figure}

As Fig.~\ref{w_K:fig} suggests, strong size effects are observed,
especially for large $K$.
In particular, the ``natural'' symmetric large-$K$ limit $w=1/2$ found with
PBC in Ref.~\cite{Vanossi06} is rarely reached using OBC: for $N=100$ the
chain is pinned to the bottom substrate ($v_{\rm cm}=0$), while for $N=15$ it moves at
$v_{\rm cm}=v_{\rm ext}$.
Figure~\ref{K1000:fig} shows the dependence of the velocity ratio $w$ on the
particle number $N$, for a fixed length ratio $r_+$ and two different
values of $r_-$, in the large-$K$ limit.
We observe that the chain is pinned to the $a_+$ substrate when $N$ occurs
to be (nearly) a multiple of the length ratio $r_-$.
In all other cases, the chain follows the $a_-$ substrate, at velocity
$v_{\rm ext}$.

This changing  behavior can be understood as follows.
For large spring stiffness $K$, the kink dynamics is suppressed, the
lubricant particles placing themselves at nearly regular distances $\simeq
a_0$.
It is energetically favorable for a chain shorter than $a_-$ to sit in a
minimum of the top potential and then stick to it.
Even if the chain is longer than $a_-$, its length is generally
not an exact multiple of the periodicity of the top substrate.
For this reason, a finite end part of the chain remains similarly trapped
in the minima of the top substrate: this is what occurs for most $N$ values
in Fig.~\ref{K1000:fig}.
On the other hand, when $N\,a_0$ is a multiple of the
$a_-$ period (i.e.\ when $N$ is close to a multiple of $r_-=a_-/a_0$),
minima and maxima of the upper potential compensate each other, so that
there is no preferential relative position of chain and top substrate.
For such special sizes, the chain remains weakly pinned to the bottom
substrate, as illustrated in Fig.~\ref{K1000:fig}.
The values $N=15$ and $N=100$ of  Fig.~\ref{w_K:fig} represent the two situations.

Other intermediate values occur for specific sizes, but the finite-size
scaling for large $N$ is obviously non-trivial.
While Fig.~\ref{w_K:fig} indicates that for small and moderate $K$,
size effects, if any, are very small, at large $K$ they affect the
dynamics substantially.
Further work is needed to understand the large-$K$ size-scaling in detail.
%

\section{Discussion and Conclusions}

We have shown that chains of finite and even small size, driven in between
two periodic sliders, move with characteristic quantized velocities, much like the
infinite-size ones do.
We find that a finite-size chain length re-adjusts in such a way
as to realize a lattice of topological solitons (kinks) commensurate
to the period of the smoother slider.
A consequence of this self-commensuration is a periodic single-particle
motion, even for incommensurate initial choices of the periods.

The likely reason behind this phenomenology is as follows.
Consider initially a single periodic slider of length ratio $r_+$
and periodic boundary conditions. The lubricant chain will then
form a regular lattice of kinks, which repel each other. When the
second slider is introduced, the lattice of kinks will be
generally incommensurate with the period $a_-$, and this brings an
irregular distribution of bunches of kinks, which on average
reconstruct the correct density of kinks. An OBC finite chain is
able to relax, by paying some extra harmonic strain, in such a way
as to enforce an optimal local density of kinks which satisfies
both periodic sliders.

The phenomena just described for a model 1D system are unique,
and it would be interesting if they could be observed in real systems.
Nested carbon nanotubes \cite{Zhang}, or confined one-dimensional nanomechanical systems
\cite{Toudic_06}, are one possible arena for the phenomena described.
Though speculative at this stage, one obvious question is what aspects of
the phenomenology just described might survive in two-dimensions (2D),
where tribological realizations, such as the sliding of two hard
crystalline faces with, e.g., an interposed graphite flake, are
conceivable.
Our results suggests that the lattice of discommensurations -- a Moir\'e
pattern-- formed by the flake on a substrate, could be dragged by the other
sliding crystal face, in such a manner that the speed of the flake as a
whole would be smaller, and quantized. This would amount to the slider
``ironing'' the kinks onward.
Dienwiebel {\it et al.}\ \cite{Dienwiebel04} demonstrated how
incommensurability may lead to virtually friction-free sliding in such a
case, but no measure was obtained for the flake relative sliding velocity.
Real substrates are, unlike our model, not rigid, subject to thermal expansion, etc.
Nevertheless the ubiquity of plateaus shown in Fig.~\ref{w_K:fig},
and their topological origin, suggests that these effects would not remove the phenomenon.

\section*{Acknowledgments}

This research was partially supported by PRRIITT (Regione Emilia Romagna), Net-Lab
``Surfaces \& Coatings for Advanced Mechanics and Nanomechanics''
(SUP\&RMAN) and by MIUR Cofin 2004023199, FIRB RBAU017S8R, and RBAU01LX5H.

%
%

\newpage

\end{document}